\documentclass[aps,preprint]{revtex4}%
\usepackage{amsfonts}
\usepackage{amsmath}
\usepackage{amssymb}
\usepackage{graphicx}%
\begin{document}
\preprint{Manuscript}
\title{Dust ion-acoustic shocks in quantum dusty pair-ion plasmas}
\author{A. P. Misra}
\email{apmisra@visva-bharati.ac.in}
\affiliation{Department of Mathematics, Siksha Bhavana, Visva-Bharati University,
Santiniketan-731 235, India.}

\begin{abstract}
The formation of dust ion-acoustic shocks (DIASs) in a four-component quantum
plasma whose constituents are electrons, both positive and negative ions and
immobile charged dust grains, is studied. The effects of both the dissipation
due to kinematic viscosity and the dispersion caused by the charge separation
as well as the quantum tunneling due to the Bohm potential are taken into
account. The propagation of small but finite amplitude dust ion-acoustic waves
(DIAWs) is governed by the Korteweg-de Vries-Burger (KdVB) equation which
exhibits both oscillatory and monotonic shocks depending not only on the
viscosity parameters $\eta_{\pm}=\mu_{\pm}\omega_{p-}/c_{s}^{2}$ (where
$\mu_{\pm}$ are the coefficients of kinematic viscosity, $\omega_{p-}$ is the
plasma frequency for negative ions and $c_{s}$ is the ion-sound speed) but
also on the quantum parameter $H$ (the ratio of the electron plasmon to the
electron Fermi energy) and the positive to negative ion density ratio $\beta.$
Large amplitude stationary shocks are recovered for a Mach number $(M)$
exceeding its critical value $(M_{c}).$ Unlike the small amplitude shocks,
quite a smaller value of $\eta_{+},\eta_{-},$ $H$ and $\beta$ may lead to the
large amplitude monotonic shock strucutres. The results could be of importance
in astrophysical and laser produced plasmas.



\end{abstract}
\date{03.06.2008}
\startpage{1}
\endpage{102}
\maketitle

{\LARGE 1.Introduction}

Physically, a pair-ion plasma is similar to an electron-positron plasma, in
which particles have the same mass and opposite charges. Negative ions are
found to be an extra component (which may occur naturally or may be injected
from external sources) in most space and laboratory plasmas [1,2]. There are a
number of works for investigating such pair ion plasmas. Recently, Kim and
Merlino [3] have discussed the conditions under which dust, when injected into
a laboratory negative ion plasma, becomes positively charged for very large
values of negative ion density $\gtrsim500$ times the electron density. Rapp
\textit{et al} [4] have discussed the possible role of negative ions in
explaining their observations of positively charged nanoparticles in the
mesosphere under nighttime conditions. Cooney \textit{et al }[5] have
investigated a two-dimensional soliton in a pair-ion plasma. Also, the role of
negative ions in a laboratory dusty plasma have been discussed by Klumov
\textit{et al} [6]. An experimental investigation of the effects of negative
ions on shock formation in a collisional Q-machine plasma has been made by Luo
\textit{et al }[7]. Moreover, It has been pointed out that such pair-ion
plasmas have potential applications in the atmosphere of D-region of the
Earth's ionosphere, the Earth's mesosphere, the solar atmosphere as well as in
the microelectronics plasma processing reactors [8]. Takeuchi \textit{et al}
[9] in their work have reported the experimental observations of ion-acoustic
shocks in an unmagnetized plasmas whose constituents are electrons as well as
positive and negative ions. They observed wave steepening of positive or
negative jumps for certain values of the positive to negative ion density
ratio $(\beta=n_{+0}/n_{-0}).$ They suggested that their experimental
observations can be well explained by means of the Korteweg de-Vries (KdV)
equation in which the nonliner coefficient $(A)$ can change its sign when a
significant fraction of negative ions is present in an electron-positive ion
plasma. Various laboratory experiments [10] have been conducted over the last
few years to study the formation of dust ion-acoustic shocks (DIASs) in dusty
plasmas. Dust ion acoustic compressional pulses have been observed to steepen
as they travel through a plasma containing negatively charged dust grains.
Theoretical models [11] have been proposed to explain the formation of small
amplitude DIASs in terms of the Korteweg-de Vries-Burger (KdVB) equation, in
which the dissipative terms comes from the dust charge perturbations [12] and
kinematic viscosity [13].

Quantum plasmas, where the finite width of the electron wave functions gives
rise to collective effects [14,15], are currently an emerging field of
research area. In view of its potential applications in micro-electronic
devices [16], nanoscale systems [17], in laser fusion plasmas [18 ], next
generation high intensity light sources [19,20] as well as in dense
astrophysical environments [21], various collective processes have been
investigated in a number of research works [e.g. see Refs.22-26].

There has also been much interests in investigating the strucure and dynamics
of shocks in quantum like systems, such as nonlinear optical fibers and
Bose-Einstein condensates [27-29]. The structure of such shocks are quite
different from the classical ones where the shocks are typically governed by
the transport processes, i.e., the viscosity and thermal conduction. Unlike
classical fluids, quantum plasmas typically exhibit dispersion due to the
quantum tunneling associated with the Bohm potential instead of dissipation
[22-26]. For this reason, even a quantum shock propagating with constant
velocity in a uniform media does not exhibit a stationary structure.
Transition from initial to compressed quantum media occurs in the form of a
train of solitons propagating with different velocities and with different
amplitudes [22-26]. Such a train of solitons also provides a non-monotonic
transition from the initial to final state of the medium. However, there are
various quantum plasma systems in which both dissipation and dispersion play
roles. Such important roles have been studied in different quantum plasma
systems for the formation of ion-acoustic shocks (where the dissipation is due
to kinematic viscosity) recently by Sahu \textit{et al} [30] and Misra
\textit{et al} [26].

It is therefore of interest to examine the effects of kinematic viscosity as
well as the quantum mechanical effects on the propagation of dust ion-acoustic
waves (DIAWs) and to show how the dispersion caused by the charge separation
as well as the density correlation due to quantum fluctuation, and the
dissipation due to kinematic viscosity play crucial roles in the formation of
shock waves instead of solitary wave solutions. We also investigate how the
presence of negative ions modify the wave structures in a multi-component
dusty quantum plasma. This is the aim of the present investigation. By using
the standard reductive perturbation technique (RPT), the small amplitude DIAWs
is described by the Korteweg-de Vries-Berger (KdVB) equation, where the Burger
term appears due to kinematic viscosity determined by both positive and
negative ions. The equation is then numerically solved to show that either
oscillatory (dispersion-dominant case ) or monotonic (dissipation-dominant
case) shock wave solutions are possible to exist depending on the
nondimensional quantum diffraction parameter $H$, the viscosity parameter
$\eta_{\pm}$ and $\beta,$ the equilibrium positive to negative ion density
ratio.\newline We have also recovered the large amplitude shock solutions for
values of the Mach number $\left(  M\right)  $ exceeding critical value
$\left(  M_{c}\right)  .$\newline

{\LARGE 2. Basic equations and small amplitude shock solutions}

We consider the propagation of DIAWs in an unmagnetized collisionless quantum
plasma composed of electrons, positive and negative ions and immobile
negatively charged dust grains. The dynamics of DIAWs in our quantum dusty
plasma is governed by the following set of hydrodynamic equations:
\begin{equation}
0=\frac{e}{m_{e}}\frac{\partial\phi}{\partial x}-\frac{1}{{m_{e}}{n_{e}}}%
\frac{\partial{p_{e}}}{\partial x}+\frac{\hbar^{2}}{2{m_{e}}^{2}}%
\frac{\partial}{\partial x}\left(  \frac{1}{\sqrt{n_{e}}}\frac{\partial
^{2}\sqrt{n_{e}}}{\partial x^{2}}\right)  , \label{eqn1}%
\end{equation}%
\begin{equation}
\frac{\partial n_{+}}{\partial t}+\frac{\partial(n_{+}u_{+})}{\partial x}=0,
\label{eqn2}%
\end{equation}

\begin{equation}
\frac{\partial u_{+}}{\partial t}+{u_{+}}\frac{\partial{u_{+}}}{\partial
x}=-\frac{Z_{+}e}{m_{+}}\frac{\partial\phi}{\partial x}+\mu_{+}\frac
{\partial^{2}u_{+}}{\partial x^{2}}, \label{eqn3}%
\end{equation}

\begin{equation}
\frac{\partial n_{-}}{\partial t}+\frac{\partial(n_{-}u_{-})}{\partial x}=0,
\label{eqn4}%
\end{equation}

\begin{equation}
\frac{\partial u_{-}}{\partial t}+{u_{-}}\frac{\partial{u_{-}}}{\partial
x}=\frac{Z_{-}e}{m_{-}}\frac{\partial\phi}{\partial x}+\mu_{-}\frac
{\partial^{2}u_{-}}{\partial x^{2}}, \label{eqn5}%
\end{equation}

\begin{equation}
\frac{\partial^{2}\phi}{\partial x^{2}}=\frac{e}{\varepsilon_{0}}\left(
n_{e}-Z_{+}n_{+}+Z_{-}n_{-}\right)  , \label{eqn6}%
\end{equation}
where $n_{\alpha},u_{\alpha},m_{\alpha}$ are respectively the density (with
equilibrium value $n_{\alpha0}$), velocity and mass for electrons
$(\alpha=e),$ positive ions $(\alpha=+)$ and negative ions $(\alpha=-)$;
$\hbar$ is the Planck's constant divided by $2\pi$; $\phi$ is the
electrostatic wave potential; $p_{e}$ is the electron pressure; $x$ and $t$
are respectively the space and time variables, and $\mu_{\pm}$ is the
coefficient of kinematic viscosity due to positive (negative) ions. At
equilibrium the overall charge neutrality condition reads
\begin{equation}
n_{e0}+Z_{-}n_{-0}+Z_{d0+}n_{d0}=Z_{+}n_{+0}, \label{eqn7}%
\end{equation}
where $Z_{\pm},$ $Z_{d0}$ are the charge states for positive (negative) ions
and dusts, $n_{d0}$ is the equilibrium dust number density. We assume that the
ions are cold, and electrons obey the following pressure law [31].
\begin{equation}
p_{e}=\frac{m_{e}v_{Fe}^{2}}{3n_{e0}^{2}}n_{e}^{3}, \label{eqn8}%
\end{equation}
where $v_{Fe}=\sqrt{2k_{B}T_{Fe}/m_{e}}$ is the electron Fermi thermal speed,
$T_{Fe}$ is the particle Fermi temperature given by $k_{B}T_{Fe}=\hbar
^{2}(3\pi^{2})^{2/3}n_{e0}^{2/3}/2m_{e}$, $k_{B}$ is the Boltzmann's constant.
Now introducing the following normalizations
\begin{equation}
x\rightarrow\omega_{p-}x/c_{s},~~{t}\rightarrow\omega_{p-}t,~~{n_{\alpha}%
}\rightarrow n_{\alpha}/n_{\alpha o},~~u_{\alpha}\rightarrow u_{\alpha}%
/c_{s},~~{\phi}\rightarrow e\phi/(k_{B}T_{Fe}), \label{eqn9}%
\end{equation}
where $\alpha=e,+,-$ and $\omega_{p\alpha}=\sqrt{n_{\alpha0}e^{2}%
/\varepsilon_{0}m_{\alpha}}$ is the $\alpha-$particle plasma frequency,
$c_{s}=\sqrt{k_{B}T_{Fe}/m_{-}}$ is the quantum ion-acoustic speed, we get the
following the normalized set of basic equations as :
\begin{equation}
0=\frac{\partial\phi}{\partial x}-2n_{e}\frac{\partial n_{e}}{\partial
x}+\frac{H^{2}}{2\mu}\frac{\partial}{\partial x}\left(  \frac{1}{\sqrt{n_{e}}%
}\frac{\partial^{2}\sqrt{n_{e}}}{\partial x^{2}}\right)  , \label{eqn10}%
\end{equation}%
\begin{equation}
\frac{\partial n_{+}}{\partial t}+\frac{\partial(n_{+}u_{+})}{\partial x}=0,
\label{eqn11}%
\end{equation}

\begin{equation}
\frac{\partial u_{+}}{\partial t}+{u_{+}}\frac{\partial{u_{+}}}{\partial
x}=-m\frac{\partial\phi}{\partial x}+\eta_{+}\frac{\partial^{2}u_{+}}{\partial
x^{2}}, \label{eqn12}%
\end{equation}

\begin{equation}
\frac{\partial n_{-}}{\partial t}+\frac{\partial(n_{-}u_{-})}{\partial x}=0,
\label{eqn13}%
\end{equation}

\begin{equation}
\frac{\partial u_{-}}{\partial t}+{u_{-}}\frac{\partial{u_{-}}}{\partial
x}=\frac{\partial\phi}{\partial x}+\eta_{-}\frac{\partial^{2}u_{-}}{\partial
x^{2}}, \label{eqn14}%
\end{equation}

\begin{equation}
\frac{\partial^{2}\phi}{\partial x^{2}}=\mu n_{e}-\beta n_{+}+n_{-},
\label{eqn15}%
\end{equation}
where $\mu=n_{e0}/Z_{-}n_{-0},$ ${\beta=Z}_{+}n_{+0}/Z_{-}n_{-0}$ connected
through the charge neutrality condition [Eq.(7)] $\mu=\beta-1-\delta$ with
$\delta=Z_{d0}n_{d0}/Z_{-}n_{-0},$ $m=m_{-}/m_{+},$ $\eta_{\pm}=\mu_{\pm
}{\omega_{p-}}/{c_{s}^{2}}$ and the nondimensional quantum parameter
$H={\hbar\omega_{pe}}/({k_{B}T_{Fe}})$ (the ratio between the electron plasmon
energy and the electron Fermi energy) proportional to quantum diffraction.
Now, integrating once the Eq. (10) with the boundary conditions \textit{viz}.
$n_{e}\rightarrow1,\frac{\partial n_{e}}{\partial x}\rightarrow0$ and
$\phi\rightarrow0$ at $\pm\infty$ we have
\begin{equation}
\phi=-1+n_{e}^{2}-\frac{H^{2}}{2\mu}\frac{1}{\sqrt{n_{e}}}\frac{\partial
^{2}\sqrt{n_{e}}}{\partial x^{2}}. \label{eqn16}%
\end{equation}
In order to investigate the propagation of small but finite amplitude DIAWs
and to derive the required governing equation in our quantum dusty pair-ion
plasma, we stretch the independent variables as $\xi=\epsilon^{1/2}(x-\lambda
t),\tau=\epsilon^{3/2}t$ with $\eta_{\pm}=\epsilon^{1/2}\eta_{\pm0}$, while
$\eta_{\pm0}$ is a finite quantity of the order of unity, while the dependent
variables are expanded as%

\begin{equation}
n_{\alpha}=1+\epsilon n_{\alpha}^{(1)}+\epsilon^{2}n_{\alpha}^{(2)}%
+\epsilon^{3}n_{\alpha}^{(3)}+\cdots, \label{eqn17}%
\end{equation}

\begin{equation}
u_{\alpha}=0+\epsilon u_{\alpha}^{(1)}+\epsilon^{2}u_{\alpha}^{(2)}%
+\epsilon^{3}u_{\alpha}^{(3)}+\cdots, \label{eqn18}%
\end{equation}

\begin{equation}
\phi=0+\epsilon\phi^{(1)}+\epsilon^{2}\phi^{(2)}+\epsilon^{3}\phi^{(3)}%
+\cdots, \label{eqn19}%
\end{equation}
where $\alpha=e,+,-$ and $\epsilon$ is a small nonzero parameter proportional
to the amplitude of the perturbation. Now, substituting the expressions from
Eqs.(17)-(19) into the Eqs.(10)-(16) and collecting the terms in different
powers of $\epsilon$, we obtain in the lowest order of $\epsilon$ the
dispersion law:%

\begin{equation}
\lambda=\pm\sqrt{\frac{2(1+m\beta)}{\mu}}. \label{eqn20}%
\end{equation}
Which shows that the DIAW can propagate outward or inward depending on the
positive or negative sign of $\lambda,$ which increases (decreases) with
$\beta$ $(\mu)$. Now, in the next higher order of $\epsilon$ we eliminate the
second order perturbed quantities from a set of equations to obtain the
required KdVB equation for DIAWs in our quantum plasma.
\begin{equation}
\frac{\partial n}{\partial\tau}+An\frac{\partial n}{\partial\xi}%
+B\frac{\partial^{3}n}{\partial\xi^{3}}+C\frac{\partial^{2}n}{\partial\xi^{2}%
}=0, \label{eqn21}%
\end{equation}
where $n\equiv n_{e}^{(1)}$ and the coefficients of the nonlinear $(A)$,
dispersive $(B)$ and dissipative $(C)$ terms are%

\begin{align}
A  &  =\sqrt{\frac{1+m\beta}{2\mu}}-\frac{3\sqrt{\mu}(1-m^{2}\beta)}%
{\sqrt{2(1+m\beta)^{3}}},B=\sqrt{\frac{1+m\beta}{128\mu^{3}}}(16-H^{2}%
),\label{eqn22}\\
C  &  =-\frac{\eta_{-0}+m\beta\eta_{+0}}{2(1+m\beta)}.\nonumber
\end{align}

We find that in Eq.(21) all the coefficients $A$ , $B$ and $C$ are modified by
the inclusion of negative ions, while $B$ and $C$ are further modified by the
effects of quantum diffraction and the effects of kinematic viscosity
respectively. In order that the DIAWs propagate with finite velocity $\lambda$
we must have $\mu(\equiv\beta-1-\delta)>0$ or $\beta>1+\delta.$ Note that in
absence of viscosity term, Eq.(21) reduces to a usual KdV equation for the
propagation of DIAWs, whereas for $H=4$ (for which $B=0)$ it reduces to a
purely Burger equation.

In order to investigate the nonlinear dynamics of DIAWs we have numerically
solved the KdVB equation (21) for different sets of parameters. The equation
(21) has a solution that can be represented as an oscillatory shock. However,
when the dissipation overwhelms the DIAW dispersion and when the dissipative
effect (kinematic viscosity) is in nice balance with the nonlinearity arising
from the nonlinear mode coupling of finite amplitude DIAWs in a dusty pair-ion
plasma, we indeed have the possibility of monotonic compressive or rarefactive
shocks. The dynamics of the latter is governed by a Burger equation. In the
numerical procedure the KdVB equation was advanced in time with a standard
fourth-order Runge-Kutta scheme with a time step of $10^{-3}$ s. The spatial
derivatives were approximated with centered second-order difference
approximations with a spatial grid spacing of $0.2$ m. The numerical solution
of Eq.(21) is shown in Figs.1 and 2 for different values of the system
parameters $\beta,$ $\eta_{\pm},$ $\delta$ and $H.$ Figure 1 shows the
oscillatory shock profiles of the electron density perturbation at the end of
$\tau=0.3$ s for $m=4$, $\beta=4,$ $\eta_{+0}=0.3,\eta_{-0}=0.2,$ $\delta=0.4$
and $H=0.2$. Even for small but non-zero values of $H$ and $\eta_{\pm0}$ we
observe a train of oscillations ahead of the shock front decaying at
$\xi\rightarrow-\infty.$ Increasing the role of quantum effects $(H=3)$ we
observe a less number of oscillations ahead of the shock wave. The amplitude
of the shock front seems to increase with the increasing values of $H$ and
$\beta,$ but decreases with $\delta.$ The oscillations decay quite fast in
case of relatively higher values of $H$ and $\beta$ and quite slow at higher
$\delta.$ Moreover, in case of large quantum effects $(H=3)$ and large
dissipative effects ($\eta_{+0}=1.0,\eta_{-0}=0.9$) and $\beta=10,\delta=0.4$
we recover monotonic transition from the oscillatory shocks (Fig.2). For
$H<3$, we again find few oscillations ahead of the shock in which first few
oscillations at the wave front will be close to solitons.The stationary
solution of Eq.(21) can also be obtained analytically. We find that $C$ always
negative, $B\lessgtr0$ according as $H\gtrless4.$ Since $H>4$ corresponds to a
lower density region (as can be found from the Fermi temperature-density
relation mentioned earlier) we consider $H<4$, so that $B>0.$ Also, since $m,$
$\beta>1,A>0.$ Thus, in our purpose $A>0,B>0$ and $C<0$. In order to find a
stationary solution of Eq.(21) we use the transformation $\zeta=\xi-U_{0}\tau$
in Eq.(21)$,$ where $U_{0}$ is the normalized velocity of the DIA shock waves
and obtain the following equation%

\begin{equation}
B\frac{d^{2}n}{d\zeta^{2}}+C\frac{dn}{d\zeta}+\frac{1}{2}An^{2}-U_{0}%
n+(U_{0}-\frac{1}{2}A)=0, \label{eqn23}%
\end{equation}
where we have imposed the boundary conditions $n\rightarrow1,dn/d\zeta
,d^{2}n/d\zeta^{2}\rightarrow0$ as $\zeta\rightarrow\infty$. Eq. (23)
describes a shock wave [32] whose velocity in the moving frame of reference is
$U_{0}.$ Since the nature of the shocks depends on the system parameters
$\beta,$ $\eta_{\pm},$ $\delta$ and $H,$ we consider the case where the
dissipation term dominates over the dispersive term. In that case, Eq.(23)
reduces to

\begin{equation}
(U_{0}-An)\frac{dn}{d\zeta}=C\frac{d^{2}n}{d\zeta^{2}}. \label{eqn24}%
\end{equation}
Which yields upon integration the following monotonic compressive shock solution%

\begin{equation}
n=\frac{U_{0}}{A}{\Large [}1-\tanh{\Large \{}-\frac{U_{0}}{2C}{\Large (}%
\xi-U_{0}\tau{\Large )\}]} \label{eqn25}%
\end{equation}
with the shock speed $U_{0},$the shock height $U_{0}/A$ and the shock
thickness $-C/U_{0}$. Since, $A$ increases with $\beta,$ and $-C$ increases
with $\beta$ and $\eta_{\pm0},$ the shock height decreases as $\beta$
increases, and the thickness increases with increasing values of $\beta$ and
$\eta_{\pm0}.$ When the dissipative effects are small, the shock will have an
oscillatory profile, whereas for large values of $\eta_{\pm0}\sim1,$ the shock
will have a monotonic behavior as is seen in the numerical solution of
Eq.(21). To determine the values of $\eta_{\pm0}$ and $\beta$ analytically
(since $C$ depends on both $\eta_{\pm0}$ and $\beta)$ corresponding to
monotonic or oscillatory shock profiles, we investigate the asymptotic
behaviors of the solution of Eq.(23) for $\zeta\rightarrow-\infty.$We
substitute $n(\zeta)=1+N(\zeta),N<<1$ in Eq.(23) and linearize to obtain%

\begin{equation}
B\frac{d^{2}N}{d\zeta^{2}}+C\frac{dN}{d\zeta}+(A-U_{0})N=0, \label{eqn26}%
\end{equation}
The solution of Eq.(26) are proportional to exp$(p\zeta),$ where%

\begin{equation}
p=-\frac{C}{2B}\pm\sqrt{\frac{C^{2}}{4B^{2}}-\frac{A-U_{0}}{B}} \label{eqn27}%
\end{equation}
It turns out that the DIA shock wave has a monotonic or oscillatory profile
according as $C^{2}\gtrless2(A-U_{0})B.$The stationary oscillatory solution of
Eq.(21) is obtained as%

\begin{equation}
n=1+D\exp\left(  -\frac{\acute{\zeta}C}{2B}\right)  \cos\left(  \acute{\zeta
}\sqrt{\frac{A-U_{0}}{B}}\right)  , \label{eqn28}%
\end{equation}
where $D$ is a constant and $\acute{\zeta}=\zeta-U_{0}\tau.$

{\LARGE 3. Large amplitude shock solutions}

We now consider large amplitude planar stationary shock. In the moving frame
of reference $\varsigma=x-Mt,$ the basic normalized equations (10)-(15) may be
integrated as%

\begin{equation}
\phi=-1+n_{e}^{2}-\frac{H^{2}}{2\mu}\frac{1}{\sqrt{n_{e}}}\frac{d^{2}%
\sqrt{n_{e}}}{d\varsigma^{2}} \label{eqn29}%
\end{equation}

\begin{equation}
n_{\pm}=\frac{M}{M-u_{\pm}} \label{eqn30}%
\end{equation}

\begin{equation}
-Mu_{\pm}+\frac{1}{2}u_{\pm}^{2}=(-\mu,1)\phi+\eta_{\pm}\frac{du_{\pm}%
}{d\varsigma} \label{eqn31}%
\end{equation}

\begin{equation}
\frac{d^{2}\phi}{d\varsigma^{2}}=\mu n_{e}-\beta n_{+}+n_{-} \label{eqn32}%
\end{equation}

Eliminating the variables $n_{e}$ and $\phi$ and using the quasi-neutrality
condition $\mu n_{e}=\beta n_{+}+n_{-}$ we obtain the following desired set of equations%

\begin{equation}
\frac{dn_{+}}{d\varsigma}=N_{+}, \label{eqn33}%
\end{equation}

\begin{equation}
\frac{dn_{-}}{d\varsigma}=\left(  f_{1}-\frac{M\eta_{+}N_{+}}{n_{+}^{2}%
}\right)  \frac{n_{-}^{2}}{\mu M\eta_{-}}, \label{eqn34}%
\end{equation}

\begin{equation}
\frac{dN_{+}}{d\varsigma}=f_{2}+f_{3}\left(  f_{1}-\frac{M\eta_{+}N_{+}}%
{n_{+}^{2}}\right)  \frac{n_{-}^{2}}{mM\eta_{-}}+f_{4}N_{+}, \label{eqn35}%
\end{equation}
where
\begin{equation}
f_{1}=-M^{2}\left[  \left(  1-\frac{1}{n_{+}}\right)  +m\left(  1-\frac
{1}{n_{-}}\right)  \right]  +\frac{M^{2}}{2}\left[  \left(  1-\frac{1}{n_{+}%
}\right)  ^{2}+m\left(  1-\frac{1}{n_{-}}\right)  ^{2}\right]  , \label{eqn36}%
\end{equation}

\begin{equation}
f_{2}=-\frac{4\mu\left(  \beta n_{+}-n_{-}\right)  }{H^{2}m(f+\beta)}\left[
\begin{array}
[c]{c}%
m+M^{2}\left(  1-\frac{1}{n_{+}}\right)  -\frac{M^{2}}{2}\left(  1-\frac
{1}{n_{+}}\right)  ^{2}+\frac{M\eta_{+}N_{+}}{n_{+}^{2}}-\\
\frac{m}{\mu^{2}}\left(  \beta n_{+}-n_{-}\right)  ^{2}-\frac{mH^{2}\left[
\beta N_{+}-\left(  f_{1}-\frac{M\eta_{+}N_{+}}{n_{+}^{2}}\right)  \frac
{n_{-}^{2}}{\mu M\eta_{-}}\right]  ^{2}}{8\mu\left(  \beta n_{+}-n_{-}\right)
^{2}}%
\end{array}
\right]  , \label{eqn37}%
\end{equation}

\begin{equation}
f_{3}=\left[  \frac{2n_{-}}{mM\eta_{-}}\left(  f_{1}-\frac{M\eta_{+}N_{+}%
}{n_{+}^{2}}\right)  -\frac{M}{n_{-}\eta_{-}}\right]  /(f+\beta),
\label{eqn38}%
\end{equation}

\begin{equation}
f_{4}=\frac{n_{-}^{2}}{m\eta_{-}n_{+}^{3}}\left(  2\eta_{+}n_{+}-M\right)
/(f+\beta),\text{ }f=\frac{\eta_{+}n_{-}^{2}}{m\eta_{-}n_{+}^{2}}
\label{eqn39}%
\end{equation}
We have numerically solved the system of equations (33)-(35) by Runge-Kutta
scheme starting from the initial conditions $n_{+}=n_{-}=1.01,N_{+}=0.001$
with a step size $\Delta\varsigma=0.001.$ It is found that the perturbations
develop into shock waves provided the Mach speed $(M)$ excceds its critical
value $M_{c}=5.$ For $m=4$, $\beta=4,\delta=0.1,$ $\eta_{+0}=0.02,\eta
_{-0}=0.01$ and $H=0.4,$ the oscillatory shock solution is shown to exist for
$M=6$ in Fig.3. As the value of $H$ decreases, the number of oscillations
ahead of the shock decreases together with the increase of the amplitude at
the upstream side. The effects of $\delta$ is to increase both the amplitude
and width of the oscillatory shock profile. Also, increasing the value of
$\beta($say$,10),$ we can clearly see a less number of peaks and troughs at
the upstream side $(\varsigma=0).$ Unlike the small amplitude case, the large
amplitude DIA monotonic shocks can exist even for comparatively small values
of $\eta_{\pm0}.$ As for example, for $M=6,$ $m=4$, $\beta=4,$ $\delta=0.1,$
$\eta_{+0}=0.2,\eta_{-0}=0.1$ and $H=0.4$ the monotonic shock profile is shown
in Fig.4. Further increasing the value of the quantum parameter $H,$ e.g., for
$H=1,$ the monotonic profile again transits into the oscillatory one (Fig.5).
We would like to stress that we have obtained both small and large amplitude
stationary and nonstationary DIA shock structures in our quantum plasma model;
the train of oscillations propagate along with the shock with the same velocity.

{\LARGE 4. Discussions and Conclusions}

In this paper, we have investigated the nonlinear propagation of DIAWs in a
four-component quantum plasma composed of electrons, positive and negative
ions and stationary negatively charged dust grains. Both the dissipative (due
to kinematic viscosity) and dispersive (due to Bohm potential) effects are
taken into consideration for the formation of DIA shock structures. Both the
stationary and nonstationary shock solutions are recovered in our quantum
plasma model. The small amplitude nonstationary and stationary DIA shock
structures are obtained numerically and analytically from the KdVB equation.
The transition from oscillatory to monotonic shocks strongly depends not only
on the quantum parameter$H$ and the viscosity parameter $\eta_{\pm}$, but also
on the positive to negative ion density ratio $\beta$ (which is mainly due to
the inclusion of negative ions) . The shock structures are also modified by
the effects of $H,\beta$ and $\delta.$ Numerical simulation reveals that the
small amplitude nonstationary shock (monotonic) transition occurs compratively
at larger values of $H(\geq3)$, $\beta(\geq8)$ and $\eta_{\pm}\sim1,$whereas
for the large amplitude shocks comparatively lower values of $H(\sim
0.4),\beta(\sim4)$ and $\eta_{\pm}(\sim0.2)$ are required. Such significant
modifications of the shock wave front structures in our quantum plasma model
could be of interest in astrophysical and laser produced plasmas. Furthermore,
it is suggested that the experiments should be designed to look for dust
ion-acoustic shocks (DIASs) and solitons in such quantum dusty pair-ion
plasmas.\newline\textbf{Acknowledgement}

This work was partially supported by the Special Assistance Program (SAP),
UGC, Government of India.

\newpage

{\Large Figure captions}

Figure 1: Small amplitude oscillatory shock profile [nonstationary solution of
Eq.(21)] for $m=4$, $\beta=4,$ $\eta_{+0}=0.3,\eta_{-0}=0.2,$ $\delta=0.4$ and
$H=0.2$.

Figure 2: Monotonic shock structure [nonstationary solution of Eq.(21)] for
$m=4$, $\beta=10,$ $\eta_{+0}=1.0,\eta_{-0}=0.9,$ $\delta=0.4$ and $H=3$.

Figure 3: Large amplitude oscillatory shock profiles for $M=6,$ $m=4$,
$\beta=4,\delta=0.1,$ $\eta_{+0}=0.02,\eta_{-0}=0.01$ and $H=0.4.$

Figure 4: Large amplitude monotonic shock transition from the oscillatory
profile in Fig.3 for $M=6,$ $m=4$, $\beta=4,$ $\delta=0.1,$ $\eta
_{+0}=0.2,\eta_{-0}=0.1$ and $H=0.4.$

Figure 5: The monotonic shock profile in Fig. 4 transits into the oscillatory
one for $H=1.$ Other parameter values remain the same as Fig.4.

\newpage

{\Large \textbf{References}}\newline

[1] H. Amemiya, B.M. Annaratone, J.E. Allen, Plasma Sources Sci.

Technol. 8 (1999) 179.

[2]\ R.N. Franklin, Plasma Sources Sci. Technol. 11 (2002) A31.

[3] S.H. Kim, R.L. Merlino, Phys. Plasmas 13 (2006) 052118.

[4] M. Rapp, J. Hedin, I. Strelnikova \textit{et al }Geophys. Res. Lett. 32

(2005) L23821.

[5] I.L. Cooney, M.T. Gavin, I. Tao, K. E. Lonngren, IEE Trans. Plasma

Sci. 19 (1991) 1259.

[6] B.A. Klumov, A.V. Ivlev, G. Morfill, JETP Lett. 78 (2003) 300.

[7] Q.-Z. Luo, N.D'Angelo, R.L. Merlino, Phys. Plasma 5 (1998) 2868.

[8] S.H. Kim, R L. Merlino, Phys. Rev. E 76 (2007) 035401.

[9] T. Takeuchi, S. Iizuka, N. Sato, Phys. Rev. Lett. 80 (1998) 77.

[10] Y. Nakamura, H. Bailung, P.K. Shukla, Phys. Rev.Lett. 83 (1999) 1602;

Q.-Z. Luo, R.L. Merlino, Phys. Plasmas 6 (1999) 3455; Q.-Z. Luo,

N.D'Angelo, R.L. Merlino, Phys. Plasmas7 (2000) 2370; Y. Nakamura,

Phys. Plasmas 9 (2002) 440; S.I. Popel, M.Y. Yu, Phys. scr.T113 (2004)

105.

[11] P.K. Shukla, Phys. Plasmas 7 (2000) 1044; A.A. Mamun, P.K.

Shukla, Phys. Plasmas 9 (2002) 1468.

{[12]} E. Tandberg-Hansen, A. G. Emslie, The Physics of Solar Flares,

Cambridge University Press, Cambridge, 1988, p.124.\newline

{[13]} P.K. Shukla, Phys. Plasmas \textbf{7} (2000) 1044.\newline

[14] D. Pines, Elementary excitations in solids, Westview Press, Boulder,

1999.

[15] D. Kremp, M. Schlanges, W.-D. Kraeft, Quantum Statistics of

Nonideal Plasmas, Springer Berlin, 2005.

{[16]} M. Marklund, P.K. Shukla, Rev. Mod. Phys. 78 (2006) 591.

[17] H.G. Craighead, Science 290 (2000) 1532.

[18] S.H. Glenzer, O.L. Landen, P. Neumayer, R.W. Lee, K. Widmann, S.W.

Pollaine, R.J. Wallace, G. Gregori, A. H\"{o}ll, T. Bornath, R. Thiele, V.

Schwarz, W.-D. Kraeft, R. Redmer, Phys. Rev. Lett. 98 (2007) 065002.

[19] M. Marklund, P.K. Shukla, Rev. Mod. Phys. 78 (2006) 591.

[20] G.A. Mourou, T. Tajima, S.V. Bulanov, Rev. Mod. Phys. 78

(2006) 309.

{[21]} Y.D. Jung, Phys. Plasmas 8 (2001) 3842; M. Opher, L O. Silva, D.E.

Dauger, V.K. Decyk, J.M. Dawson, \textit{ibid} 8 (2001) 2544; G, Chabrier, E.

Douchin, Y. Potekhin, J. Phys.: Condens. Matter 14 (2002) 9133, A.Y.

Potekhin, G. Chabrier, D. Lai, W.C. G. Ho, M. van Adelsberg, J. Phys. A:

Math. Gen. 39 (2006) 4453; A.K. Harding, D. Lai, Rep. Prog. Phys. 69 (2006)

2631. \newline

{[22]} P.K. Shukla, B. Eliasson, Phys. Rev. Lett. 96 (2006) 245001;
\textit{ibid.}

99 (2007) 096401; D. Shaikh, P.K. Shukla, Phys. Rev. Lett. 99 (2007)
12502.\newline

{[23]} A.P. Misra, P.K. Shukla, C. Bhowmik, Phys. Plasmas 14 (2007) 082309;

A. P. Misra, P.K. Shukla, \textit{ibid} 14 (2007) 082312; \newline

\textit{ibid} 15 (2008) 052105; C. Bhowmik, A. P. Misra, P. K. Shukla,
\textit{ibid} 14

(2007) 122107.

[24] P.K. Shukla, S. Ali, L. Stenflo, M. Marklund, Phys. Plasmas 13 (2006)

112111.\newline

{[25]} L.G. Garcia, F. Haas, L.P.L. de Oliveira, J. Goedert, Phys. Plasmas

1\textbf{2} (2005) 12302; F. Haas, G. Manfredi, J. Goedert, Phys. Rev. E 64 (2001)

026413; Braz. J. Phys. 33 (2003) 128. \newline

[26] K. Roy, A.P. Misra, P. Chatterjee, Phys. Plasmas 15 (2008) 032310.

[27] W. Wan, S. Jia, J. Fleischer, Nature Phys. 3 (2007) 46.

[28] A. Kamchatnov, R. Kraenkel, B. Umarov, Phys. Rev. E 66 (2002)

036609; A. Kamchatnov, A. Gammal, R. Kraenkel, Phys. Rev. A 69 (2004)

063605.

[29] B. Damski, Phys. Rev. A 69 (2004) 043610.

[30] B. Sahu, R. Roychoudhury, Phys. Plasmas 14 (2007) 072310.

{[31]} L.D. Landau, E.M. Lifshitz, Statistical Physics, Oxford University

press, Oxford, 1998.\newline

[32] V.I. Karpman, Nonlinear waves in dispersive media, Oxford:

pergamon, 1975.

\bigskip

\bigskip

\bigskip

\bigskip

\bigskip

\end{document}